# Large-Angle Cosmic Microwave Background Anisotropies in an Open Universe[†]

Marc Kamionkowski[‡]

*School of Natural Sciences, Institute for Advanced Study, Princeton, NJ 08540*

and

David N. Spergel[*]

*Princeton University Observatory, Princeton, NJ 08544*

## ABSTRACT

If the Universe is open, scales larger than the curvature scale may be probed by observation of large-angle fluctuations in the cosmic microwave background (CMB). We consider primordial adiabatic perturbations and discuss power spectra that are power laws in volume, wavelength, and eigenvalue of the Laplace operator. Such spectra may have arisen if, for example, the Universe underwent a period of "frustrated" inflation. The resulting large-angle anisotropies of the CMB are computed. The amplitude generally increases as $\Omega$ is decreased, but decreases as $h$ is increased. Interestingly enough, for all three *ansatzes*, anisotropies on angular scales larger than the curvature scale are suppressed relative to the anisotropies on scales smaller than the curvature scale, but cosmic variance makes discrimination between various models difficult. Models with $0.2 \lesssim \Omega h \lesssim 0.3$ appear compatible with CMB fluctuations detected by COBE and the Tenerife experiment and with the *amplitude and spectrum* of fluctuations of galaxy counts in the APM, CfA and QDOT surveys. COBE normalization for these models yields $\sigma_8 \simeq 0.5 - 0.7$. Models with smaller values of $\Omega h$ when normalized to COBE require bias factors in excess of 2 to be compatible with the observed galaxy counts on the $8h^{-1}$ Mpc scale. Requiring that the age of the universe exceed 10 Gyr implies that $\Omega \gtrsim 0.25$, while requiring that the age exceed 13 Gyr implies that $\Omega \gtrsim 0.35$. Unlike in the flat-Universe case where the anisotropy comes only from the last-scattering term in the Sachs-Wolfe formula, large-angle anisotropies come primarily from the decay of potential fluctuations at $z \lesssim \Omega^{-1}$. Thus, if the Universe is open, COBE has been detecting temperature fluctuations produced at moderate redshift rather than at $z \sim 1300$.

---

# 1. INTRODUCTION

The recent discovery of anisotropies in the cosmic microwave background (CMB) (Smoot et al. 1992) is of tremendous importance for cosmology (for a recent review, see White, Scott, & Silk 1993). It confirms our basic notions that the large-scale structure in the Universe grew via gravitational instability from tiny density perturbations in the early Universe. Hundreds of calculations of the large-angle anisotropies expected from models for generation of the primordial density perturbations have appeared in the literature. Inflationary models and topological-defect models with hot or cold dark matter, or both, and the contribution from gravitational waves have been considered extensively. In addition, the effects of varying uncertain cosmological parameters such as the baryon density, Hubble parameter, and cosmological constant have also been studied (Bond et al. 1993).

However, with the notable exception of literally a handful of papers, all the careful calculations of large-angle anisotropies apply only to a flat Universe. Although there are plenty of good theoretical arguments for a flat Universe, many observations are at odds with $\Omega = 1$. Concordance of high values for the Hubble parameter (Jacoby et al. 1992) with conservative lower bounds on the age of the Universe is possible only in an open (or cosmological-constant) Universe. In addition, numerous dynamical measures (Hughes 1989; Regos & Geller 1989; White 1992; Fisher et al. 1993b; Davis & Peebles 1983) give $\Omega < 1$. As a matter of fact, there is no convincing observational evidence for a flat Universe, although some surveys of large-scale structure do yield values of $\Omega$ that are consistent with unity. Clearly, determination of the geometry of the Universe is one of the most pressing and challenging problems in cosmology.

The effects of geometry are manifest most strongly at redshifts $z \gg 1$, so large-angle CMB anisotropies provide perhaps the clearest probe of the geometry of spacetime. If the Universe is open, the curvature scale at the time of last scattering subtends an angle

$$\theta_{\text{curv}} \simeq \sqrt{\frac{\Omega}{1-\Omega}},$$
$$\simeq \begin{cases} 20° & \text{for } \Omega \simeq 0.1, \\ 40° & \text{for } \Omega \simeq 0.3, \end{cases} \quad (1.1)$$

so when we look at anisotropies on larger scales, we are looking directly at scales larger than the curvature scale.

Roughly speaking, the $l$th multipole moment measures power on an angular scale $\pi/l$. Thus, if the angle subtended by the curvature scale at last scattering is given by Eq. (1.1), then multipole moments $l \lesssim l_c \sim \pi/\theta_{\text{curv}}$ (e.g., $l_{\text{curv}} \simeq 4-5$ for $\Omega \simeq 0.3$, and $l_{\text{curv}} \simeq 9$ for $\Omega \simeq 0.1$) probe scales larger than the curvature scale.

Calculation of the large-angle anisotropies in a flat matter-dominated universe is straightforward and relatively simple. Given a power spectrum of density perturbations $P(k) \propto k^n$ with a given spectral index $n$, normalized to observations on scales probed by large-scale surveys, the power spectrum of curvature perturbations $\Phi$ is obtained through the Poisson equation. The temperature anisotropies are simply related to the curvature perturbation through the Sachs-Wolfe formula, $\Delta T/T = (1/3)\Phi$, and an expression for the lowest multipole moments can be given analytically.

In an open universe, the calculation is far more complicated. The spectrum of curvature perturbations at the current epoch is obtained from the spectrum of density perturbations through the appropriate generalization of the Laplace equation (Bardeen 1980); however, unlike in the flat-universe case where the curvature perturbation is time-independent, the curvature perturbation in an open universe decays with time, and this time dependence must be taken into account. Since the curvature perturbation is time dependent, the line-of-sight integral in the Sachs-Wolfe formula must be considered in addition to the last-scattering term. Still, for large-angle anisotropies, the most fundamental difference from the case of a flat universe is that Fourier analysis breaks down on scales larger than the curvature scale. Harmonic analysis on a space of constant negative curvature must be used. Although the appropriate radial functions which generalize spherical Bessel functions are known (Wilson 1983; Abbott & Schaeffer 1986), the expressions are quite complicated. Consequently, there are no simple analytic expressions for the lowest multipole moments, and they must be evaluated numerically.

Yet another difference is that if the Universe is flat, there is a well-developed class of theories (inflationary theories) that lead to the one-parameter family of primordial power-law spectra of adiabatic perturbations usually considered (Davis et al. 1992). Similar ideas have been presented for open universes (Gott 1982; Lyth & Stewart 1990; Ratra & Peebles 1993; Kamionkowski & Liddle 1993; Caldwell 1993), although at this point they have not been studied in nearly as much detail. It should be noted that even if the Universe is open, it may have undergone a period of inflation which, for some reason, ended abruptly. Although it is not clear whether such a period of "frustrated" inflation could solve the horizon problem, it could indeed provide a causal mechanism for producing primordial adiabatic density perturbations. It is also useful to recall that before the advent of inflationary theories, power-law spectra of primordial density perturbations, especially scale-invariant spectra (Harrison 1970; Peebles & Yu 1970; Zeldovich 1972), were proposed based on simple physical arguments.

In this paper, we make several plausible *ansatzes* for the primordial power spectrum, and investigate their consequences for the CMB. We do *not* discuss causal mechanisms for producing these spectra. This is left for future research. Still, if a period of frustrated inflation occurred in an open universe, the power spectrum on scales smaller than the curvature scale will resemble a standard scale-invariant spectrum, although it is still not clear what the spectrum will be on larger scales. We will see, however, that our results for CMB anisotropies are relatively insensitive to the spectrum assumed on scales larger than the curvature scale. Therefore, our results should provide a fairly good idea of what CMB anisotropies might look like if a period of frustrated inflation resulted in an open universe.

In the following Section, we briefly review the geometry of an open universe and give a simple review of the harmonic analysis on a space of constant negative curvature needed here. In Section 3, we discuss correlation functions in an open universe and present our *ansatzes* for the primordial power spectra. In Section 4, we obtain the spectrum of curvature perturbations and present formulas for the multipole moments, and in Section 5, we present some analytic approximations for CMB anisotropies on scales smaller than the curvature scale yet larger than the horizon at last scattering. We isolate the various effects of geometry, time-dependence of the potential, and normalization to observed power on small scales on the amplitude of large-angle CMB anisotropies. In Section 6, we describe the numerical results of our calculations, and in the final Section we summarize and make some concluding remarks.

## 2. GEOMETRY OF THE OPEN UNIVERSE

First let us review some kinematics and dynamics. In an open universe the metric is

$$ds^2 = dt^2 - R^2(t)[d\chi^2 + \sinh^2\chi(d\theta^2 + \sin^2\theta d\phi^2)], \qquad (2.1)$$

where we have taken $K = -1$, and $R(t)$ is the scale factor. In these variables, $\chi$ is the comoving distance in units where the curvature scale is $\chi = 1$. In physical units, the comoving distance is $\chi_{\rm phys} = H_0^{-1}(1-\Omega)^{-1/2}\chi$, where $H_0 = 100h$ km sec$^{-1}$ Mpc$^{-1}$ is the present value of the Hubble parameter (throughout, the subscript "0" denotes today). The Friedmann equation is

$$H^2 = \left(\frac{\dot R}{R}\right)^2 = \frac{8\pi G\rho}{3} + \frac{1}{R^2}, \qquad (2.2)$$

where $\rho$ is the density of the Universe, and the dot denotes a derivative with respect to time $t$. In terms of conformal time, $\eta = \int dt/R(t)$, $R(\eta) \propto (\cosh\eta - 1)$.

The relevant harmonic theory in a space of constant negative curvature (which generalizes Fourier theory) has been discussed in Wilson (1983) and Abbott & Schaeffer (1986). Here we give a brief discussion with the aim of developing an intuition for the effects of geometry in an open universe. As an introduction, consider *flat* space, where regular spherically symmetric solutions to the Helmholtz equation, $(\triangle + q^2)Q = 0$, are $Q(\chi) = \sin(q\chi)/(q\chi)$, and $\chi$ is again the comoving distance. These solutions oscillate with wavelength $2\pi/q$, so we identify the eigenvalue $q$ of the Laplace operator with the comoving wavenumber $k$.

In an open universe, the action of the Laplace operator on an arbitrary function $Q$ is

$$\triangle Q = \frac{1}{\sinh^2 \chi} \left[ \frac{\partial}{\partial \chi} \left( \sinh^2 \chi \frac{\partial Q}{\partial \chi} \right) + \frac{1}{\sin^2 \theta} \frac{\partial}{\partial \theta} \left( \sin \theta \frac{\partial Q}{\partial \theta} \right) + \frac{1}{\sin^2 \theta} \frac{\partial^2 Q}{\partial \theta^2} \right], \quad (2.3)$$

and regular spherically symmetric solutions to the Helmholtz equation are

$$Q(\chi) = \frac{\sin(\sqrt{q^2 - 1}\chi)}{\sqrt{q^2 - 1} \sinh \chi}. \quad (2.4)$$

If $q > 1$ the solutions oscillate. The wavenumber $k$ (defined to be $2\pi$ divided by the comoving wavelength) of the perturbation is *not* coincident with the eigenvalue $q$ of the Laplace operator, but they are related by $k^2 = q^2 - 1$. (We should point out that our $k$ is the same as $\beta$ in Abbott & Schaeffer (1986) and $\nu$ in Wilson (1983) and Gouda, Sugiyama, & Sasaki (1991); what we call $q$ is referred to as $k$ in all these papers. We introduce our notation since $k$ is so commonly associate with wavenumber.) Note that in units of Mpc$^{-1}$, the physical wavenumber is $k_{\text{phys}} = k H_0 \sqrt{1-\Omega}$. More generally, the eigenfunctions of the Laplace operator are $X_k^l(\chi) Y_l^m(\theta, \phi)$, and these functions form a complete set for $1 < q < \infty$. In fact, although solutions to the Helmholtz equation with $q < 1$ do exist, they can be written in terms of those with $q > 1$. The radial eigenfunctions are given by (see, e.g., Wilson 1983)

$$X_k^l(\chi) = (-1)^{l+1} N_l^{-1}(k)(k^2+1)^{l/2} \sinh^l \chi \frac{d^{l+1}(\cos k\chi)}{d(\cosh \chi)^{l+1}}, \quad (2.5)$$

where $N_l(k) = k^2(k^2+1)...(k^2+l^2)$. The normalization is chosen so that if the limit $\Omega \to 1$ is taken with $k_{\text{phys}}$ and $\chi_{\text{phys}}$ fixed, the radial eigenfunctions become spherical Bessel functions, as they should.

In a flat universe, the amplitude of the radial eigenfunctions decreases at large distances as inverse powers of $\chi$, whereas in an open universe, the amplitude drops exponentially. This is due to the fact that the volume increases exponentially with distances at distances large compared with the curvature scale. As a simple physical example of the effects of curvature in an open universe, consider Gauss' law (Callan & Wilczek 1990). In a flat universe, the electric field due to a point charge at a distance $\chi$ from the charge is $\propto \chi^{-2}$, but in an open universe, the electric field falls as $\sinh^{-2} \chi$. Similarly, the physics that leads to correlation functions that fall as power laws in a flat universe should lead to correlation functions that fall exponentially in an open universe.

## 3. CORRELATION FUNCTIONS AND POWER SPECTRA

A power law in the wavenumber $k$ in a flat universe is equivalent to a power law in volume, and to a power law in the eigenvalue of the Laplace operator. In an open universe, power laws in these three quantities are distinct, so we will consider all three. It is clear what power spectra in wavenumber $k$ and eigenvalue $q = \sqrt{k^2 + 1}$ are. Now we discuss power laws in volume.

If $\langle (\delta N/N)^2 \rangle_V$ is the square of the variance in the number of objects in a volume $V$, then the correlation function, the excess probability over random of finding an object within a distance $\chi$ of a given object, is (Peebles 1980)

$$\xi(\chi) = \frac{d}{dV} \left[ V(\chi) \left\langle \left( \frac{\delta N}{N} \right)^2 \right\rangle_{V(\chi)} \right], \quad (3.1)$$

where $V(\chi)$ is the volume enclosed within a radius $\chi$. In a flat universe, the power spectrum $P(k)$ and correlation function for a given distribution are related by

$$\xi(\chi) = \frac{1}{2\pi^2} \int k^2 \, dk \, P(k) \frac{\sin k\chi}{k\chi}, \quad (3.2)$$

and

$$P(k) = 4\pi \int \chi^2 \, d\chi \, \xi(\chi) \frac{\sin k\chi}{k\chi}, \quad (3.3)$$

so, for example, if $P(k) \propto k^n$ (for example, $n = 0$ for Poisson fluctuations and $n = 1$ for a scale-invariant spectrum), then $\xi(\chi) \propto \chi^{-n-3}$.

In an open universe, Eq. (3.1) still holds, but now (Wilson 1983)

$$\xi(\chi) = \frac{1}{2\pi^2} \int k^2 \, dk \, P(k) \frac{\sin k\chi}{k \sinh \chi}, \quad (3.4)$$

and

$$P(k) = 4\pi \int \sinh^2 \chi \, d\chi \, \xi(\chi) \, \frac{\sin k\chi}{k \sinh \chi}, \quad (3.5)$$

From Eq. (3.1), it is clear that in a flat universe,

$$\left\langle \left(\frac{\delta N}{N}\right)^2 \right\rangle_{V(\chi)} \sim \frac{1}{\langle N \rangle_V} \sim \frac{1}{V^\alpha}, \quad (3.6)$$

where $\alpha = 1 + n/3$, and $\langle N \rangle_V$ is the average number of objects in a volume $V$. Now, generalizing this volume scaling to an open universe, we find

$$\xi(\chi) \sim V^{-\alpha} \sim \frac{1}{[\sinh(2\chi) - 2\chi]^\alpha}, \quad (3.7)$$

since the volume enclosed by a sphere of radius of $\chi$ in a space of constant negative curvature is $V(\chi) = \pi[\sinh(2\chi) - 2\chi]$. Note that this reduces to the flat-space result for scales smaller than the curvature scale, $\chi \ll 1$. Then, the power spectrum which follows from a power-law correlation function in volume in an open universe is

$$P(k) = 4\pi B \int \sinh^2 \chi \, d\chi \, \frac{\sin k\chi}{k \sinh \chi} \frac{1}{[\sinh(2\chi) - 2\chi]^\alpha}, \quad (3.8)$$

where $B$ is a normalization constant. The integral in Eq. (3.8) is formally divergent at the $\chi \to 0$ limit for $\alpha \geq 0$ (the same occurs in flat space; Peebles 1980), but realistically, the correlation function becomes zero at some small distance. So, to evaluate the integral, a cutoff at small $\chi$ can be introduced, for example. Instead, it is easier to write $P(k) = \int_0^k (dP/dk) dk$. Doing so, we find

$$\frac{dP}{dk} \simeq \begin{cases} -\frac{1}{3} k^2 B I_0 & \text{for } k \to 0, \\ -B I_\infty k^{3\alpha - 4} & \text{for } k \to \infty. \end{cases} \quad (3.9)$$

where

$$I_0 = \int_0^\infty \frac{\chi^3 \sinh \chi}{[\sinh(2\chi) - 2\chi]^\alpha} \, d\chi \quad (3.10)$$

and

$$I_\infty = \left(\frac{3}{4}\right)^\alpha \sqrt{\frac{\pi}{2}} 2^{\frac{5}{2} - 3\alpha} \frac{\Gamma\left(\frac{5}{2} - \frac{3\alpha}{2}\right)}{\Gamma\left(\frac{3\alpha}{2}\right)}. \quad (3.11)$$

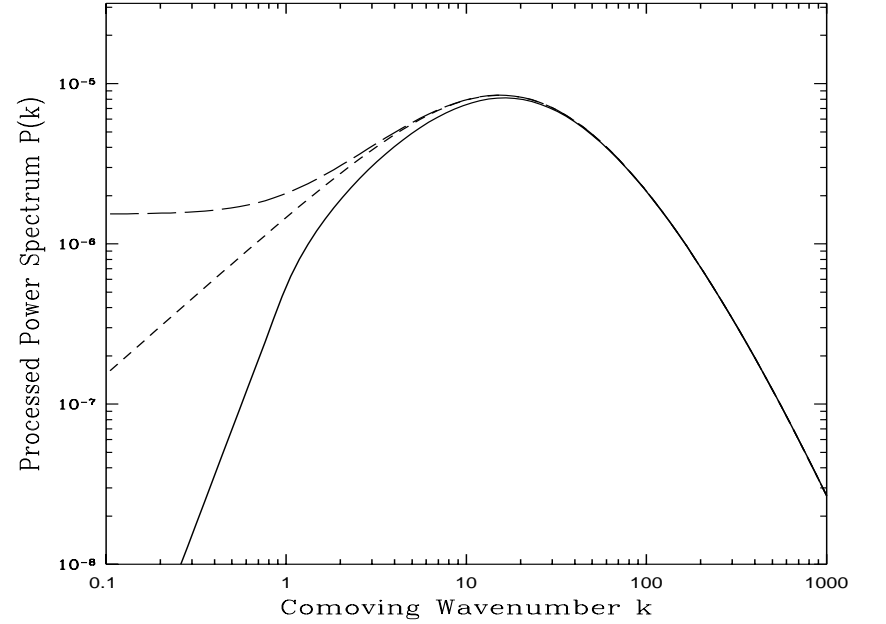

FIG. 1. Processed power spectra for primordial spectra that are power laws in volume (solid curve), wavelength (short-dash curve), and eigenvalue of the Laplace operator (long-dash curve). In all cases, the power-law index is $n = 1$.

Then, by integrating, the power spectrum can be written

$$P(k) = \begin{cases} -\frac{1}{9} B I_0 k^3 & \text{for } k \leq k_c, \\ -\frac{1}{9} B I_0 k_c^3 - B I_\infty \left(\frac{1}{3\alpha - 3}\right) \left[k^{3\alpha - 3} - k_c^{3\alpha - 3}\right], & \text{for } k > k_c, \end{cases} \quad (3.12)$$

to a good approximation where

$$k_c = \left(\frac{3 I_\infty}{I_0}\right)^{\frac{1}{6 - 3\alpha}}. \quad (3.13)$$

Until now we have been discussing our *ansatz* for the primordial power spectrum. To relate this to the power spectrum today on scales measured by galaxy surveys, the power spectrum should be multiplied by the square of a transfer function, $T^2(k)$. If we assume that the baryonic contribution to the mass density is negligible, and assume that the dark matter is cold, we can use the transfer function of Bond and Efstathiou (1984),

$$T(k) = [1 + (ak_{\text{phys}} + (bk_{\text{phys}})^{3/2} + (ck_{\text{phys}})^2)^\nu]^{-1/\nu}, \qquad (3.14)$$

where

$$a = 6.4(\Omega h^2)^{-1}\,\text{Mpc}, \quad b = 3.0(\Omega h^2)^{-1}\,\text{Mpc},$$
$$c = 1.7(\Omega h^2)^{-1}\,\text{Mpc}, \quad \nu = 1.13. \qquad (3.15)$$

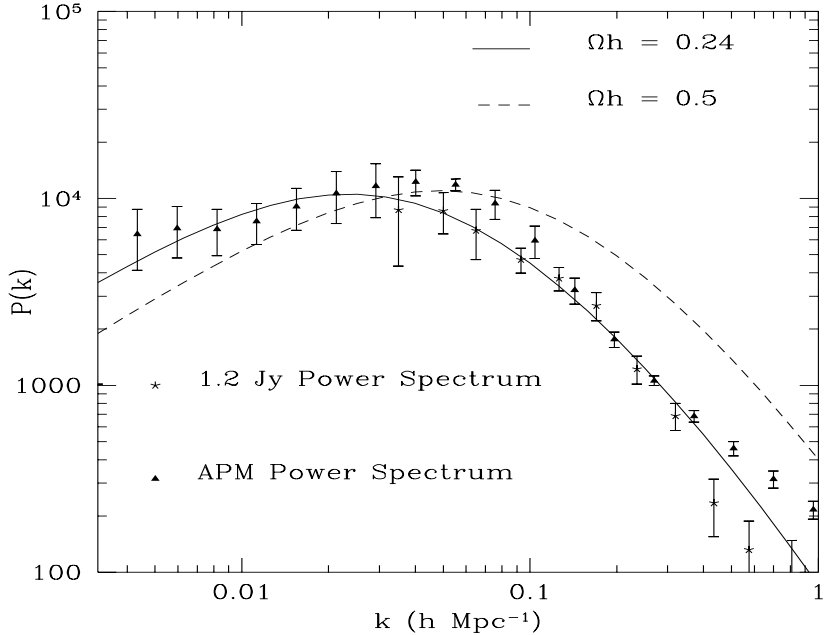

FIG. 2. Processed scale-invariant spectrum with $\Omega h = 0.24$ and $\Omega h = 0.5$.

Doing so, the power spectrum can then be normalized to $\sigma_8$, the variance of the galaxy distribution on scales of $x_f = 8h^{-1}$ Mpc (as determined from the 14.5-mag CfA survey; Davis & Peebles 1983),

$$\sigma_8^2 = \frac{1}{b^2} = \frac{1}{2\pi^2}\int k^2\,dk\, P(k)T^2(k_{\text{phys}})W^2(k_{\text{phys}}x_f), \qquad (3.16)$$

where $b$ is the bias on $8h^{-1}$ Mpc (i.e. $\sigma_8^{\text{mass}} \equiv \sigma_8/b$ where $\sigma_8^{\text{mass}}$ is the variance of the *mass* distribution on scales of $8h^{-1}$ Mpc), and $W(x) = 3[\sin x - x\cos x]/x^3$ is the top-hat window function.

In Fig. 1 are plotted the processed power spectra for primordial spectra that are power laws in volume, wavelength, and eigenvalue of the Laplace operator. On scales smaller than the curvature scale ($k \gtrsim 1$), where large-scale surveys probe the power spectrum, the spectra are indistinguishable, but on scales greater than the curvature scale ($k \lesssim 1$), those probed by large-angle anisotropies, they can be quite different. The volume scaling has the least power on large scales (it falls as $k^3$ as $k \to 0$); the wavenumber scaling goes like $k$ at small $k$; and the eigenvalue scaling has power on arbitrarily large scales (since the eigenvalue $q \to 1$ as $k \to 0$).

In Fig. 2 we plot a processed scale-invariant spectrum with $\Omega h = 0.24$ and with $\Omega h = 0.5$. The data points are from the 1.2 Jansky IRAS (Fisher et al. 1993a) and APM (Baugh & Efstathiou 1993) surveys. On these scales, power laws in $k$, $q$, and volume are all similar. Fig. 2 suggests that the large-scale power is better fit by $\Omega h \sim 0.24$ than by $\Omega h \sim 0.5$ (see also Peacock & Dodds 1993). Recall also that the theoretical curves were obtained with the assumption that the baryonic contribution to the mass density of the Universe was negligible. If the baryon density is increased, the shape of the spectrum may change, and could perhaps account for the apparent plateau in the APM data at the largest scales (Kamionkowski & Spergel 1993).

## 4. CURVATURE PERTURBATIONS AND CMB ANISOTROPIES

In an open universe, the perturbation in the density on a given scale is related to the curvature perturbation $\Phi_k$ on a scale $k$ by (Bardeen 1980)

$$(k^2 + 4)\Phi_k = 4\pi G\rho R^2 \left(\frac{\delta\rho}{\rho}\right)_k. \qquad (4.1)$$

Using the Friedmann equation, Eq. (2.2), the curvature power spectrum may be written

$$\langle|\Phi_k^0|^2\rangle = \left[\frac{3\Omega}{2(1-\Omega)(k^2+4)}\right]^2 P(k)T^2(k), \qquad (4.2)$$

where the superscript "0" refers to the value of the curvature perturbation today. Unlike in a flat universe, where the curvature perturbation remains constant in time, in an open universe the curvature decays with time. The time dependence of $\Phi_k$ is

$$\Phi_k(\eta) = \Phi_k^0 \frac{F(\eta)}{F(\eta_0)}, \tag{4.3}$$

where (Mukhanov, Brandenberger, & Feldman 1992)

$$F(\eta) = 5\frac{\sinh^2 \eta - 3\eta \sinh \eta + 4\cosh \eta - 4}{(\cosh \eta - 1)^3}. \tag{4.4}$$

The temperature autocorrelation function $C(\xi)$ as a function of angle $\xi$ can be expanded in terms of multipole moments

$$C(\xi) = \frac{1}{4\pi}\Sigma_l(2l+1)C_l P_l(\cos\xi), \tag{4.5}$$

where the sum is over all multipole moments $C_l$, and $P_l$ are the Legendre polynomials. Given the curvature power spectrum and the time dependence of the curvature perturbation, the contribution of a given mode $k$ to the $l$th multipole moment is (Gouda, Sugiyama, & Sasaki 1991; Spergel 1993)

$$\begin{aligned}\theta_l(k) &= \frac{1}{3}\Phi_k(\eta_{ls})X_k^l(\eta_0 - \eta_{ls}) + 2\int_{\eta_{ls}}^{\eta_0} \frac{d\Phi_k(\tilde{\eta})}{d\eta}X_k^l(\eta_0 - \tilde{\eta})\,d\tilde{\eta}\\ &= \Phi_k(\eta = 0)\left[\frac{1}{3}F(\eta_{ls})X_k^l(\eta_0 - \eta_{ls}) + 2\int_{\eta_{ls}}^{\eta_0}\frac{dF}{d\eta}X_k^l(\eta_0 - \tilde{\eta})\,d\tilde{\eta}\right]\\ &\equiv \Phi_k(\eta = 0)\tilde{\theta}_l(k),\end{aligned} \tag{4.6}$$

where the subscript "$ls$" denotes the value at the surface of last scattering. Note that Eq. (4.6) is valid only on scales large compared with horizon at at the surface of last scattering (multipole moments $l \lesssim 200\Omega^{-1}$). Note also that Eq. (4.6) is valid only for primordial adiabatic perturbations. Additional terms must be added if primordial entropy perturbations are to be considered (Gouda, Sugiyama, & Sasaki 1991). The CMB anisotropy is due to curvature perturbations at the surface of last scattering [the first term in Eq. (4.6)], and due to the time-evolution of perturbations along the line of sight [the second term in Eq. (4.6)]. In a flat universe, $F(\eta)$ is constant, so there is no contribution to the CMB anisotropy from the line-of-sight integral, and $\tilde{\theta}_l(k) \to j_l[k(\eta_0 - \eta_{ls})]$.

If one assumes standard recombination, then in an open universe the CMB fluctuations on small angular scales are generally too small to account for the observed large-scale structure. Reionization is often invoked to lower the amplitude of CMB anisotropies on smaller angular scales. The possible effects of reionization on large angular scales can be described by including a visibility function $\mathcal{V}(\eta)$ (Spergel 1993)

$$\tilde{\theta}_l = \int_0^{\eta_o}\left[\frac{1}{3}F(\tilde{\eta})\frac{d\mathcal{V}(\tilde{\eta})}{d\eta} + 2\frac{d\Phi_k(\tilde{\eta})}{d\eta}\mathcal{V}(\tilde{\eta})\right]X_k^l(\eta_0 - \tilde{\eta})d\tilde{\eta}. \tag{4.7}$$

We recover Eq. (4.6) by taking $\mathcal{V}(\eta)$ to be a step function. The factor of $1/3$ in the first term in Eq. (4.7) is valid only if the CMB photons scatter last at redshifts $z \gg \Omega^{-1}$, before the Universe becomes curvature dominated. In practice, unless recombination takes place at very late times ($z \lesssim \Omega^{-1}$), reionization does little to damp CMB anisotropies on angular scales larger than that subtended by the horizon at decoupling.

By integrating over all modes, we obtain the multipole moments (Spergel 1993),

$$C_l = \frac{4\pi}{2\pi^2}\int d\ln k\, k\frac{N_l(k)}{(k^2 + 1)^l}|\tilde{\theta}_l(k)|^2\left[\frac{3\Omega}{2(1-\Omega)(k^2+4)F(\eta_0)}\right]^2 P(k)T^2(k). \tag{4.8}$$

This expression is considerably more complicated than its flat-space analogue. In a flat universe, the measure is $k^3$; here it is $kN_l(k)/(k^2+1)^l$. The factor of $(k^2+4)^2$ in the denominator of the integrand reduces to $k^4$ in a flat universe. The functions $\tilde{\theta}_l(k)$ are simply spherical Bessel functions in a flat Universe, and the power spectrum $P(k)$ is taken to be a power law. In addition, the time-dependence of the curvature perturbation appears explicitly in the denominator of the integrand. Thus, unlike in the flat-space case where the lowest multipole moments can be evaluated analytically, we must evaluate Eq. (4.8) numerically.

As an aside, it was pointed out by Grishchuk & Zel'dovich (1978) that perturbations on scales larger than the horizon can in fact contribute to CMB anisotropies, and especially, to the quadrupole moment. It has been argued that the Grishchuk-Zel'dovich effect can be used to show that the Universe is smooth not only out to the current horizon, but to distance scales at least two orders of magnitude larger (Turner 1991; Kashlinsky 1993; Frieman, Kashlinsky, & Tkachev 1993). It is also argued that this implies that if the Universe inflated, it must be very close to flat today. However, the Grishchuk-Zel'dovich effect has only been worked out in a flat Universe. If the Universe is open, $\Omega \sim 0.3$ say, then the curvature scale is comparable to the current horizon, and the flat-space result does not apply. Here we simply point out that the proper contribution of super-horizon sized modes to the CMB quadrupole in an open universe may be obtained by taking the $k \to 0$ limit of Eqs. (4.8) and (4.6).

## 5. ANALYTIC APPROXIMATIONS

Before we discuss the numerical results for the multipole moments, we will present some useful analytic approximations for multipole moments, $l \gg l_{\rm curv}$, which probe scales smaller than the curvature scale. Although they may not provide a great degree of precision, they will serve to illustrate the various sources of difference between large-angle CMB anisotropies in an open universe and those in a flat universe. For simplicity, we consider only a scale-invariant ($n = 1$) spectrum. The results for other power-law indices should be qualitatively similar.

On scales smaller than the curvature scale, microwave anisotropies come predominantly from perturbations with wavenumbers $k \gg 1$. For $k \gg 1$ the power spectrum is $P(k) \propto k^n$, and if we restrict ourselves to multipole moments which probe scales smaller than the curvature scale, yet larger than the scale of the horizon at last scattering, then $T(k) \simeq 1$. Then, Eq. (4.8) becomes

$$C_l \simeq \frac{9\pi\Omega^2(\sigma_8^{\rm mass})^2}{[F(\eta_0)]^2 \mathcal{I}(\Omega,h)} \int [\tilde{\theta}_l(k)]^2 \frac{dk}{k} \qquad \text{for } l \gg l_{\rm curv}. \quad (5.1)$$

The quantity $F(\eta_0)$ is a function of $\Omega$ only and is plotted in Fig. 3. It rises monotonically with $\Omega$, but note that $\Omega/F(\eta_0) < 1$ for $\Omega < 1$. The function $\mathcal{I}(\Omega,h)$ comes from the normalization to $\sigma_8 = 1$ [c.f. Eq. (3.16)] and is plotted in Fig. 4 as a function of $\Omega$ for various values of $h$. Note that $\mathcal{I}(\Omega,h)$ decreases as either $h$ or $\Omega$ are decreased. Also, note that $\mathcal{I}$ changes by roughly an order of magnitude over the range $0.4 < h < 1.0$ and by roughly two orders of magnitude over the range $0.1 < \Omega < 1.0$. We also explicitly include the dependence on $\sigma_8^{\rm mass}$.

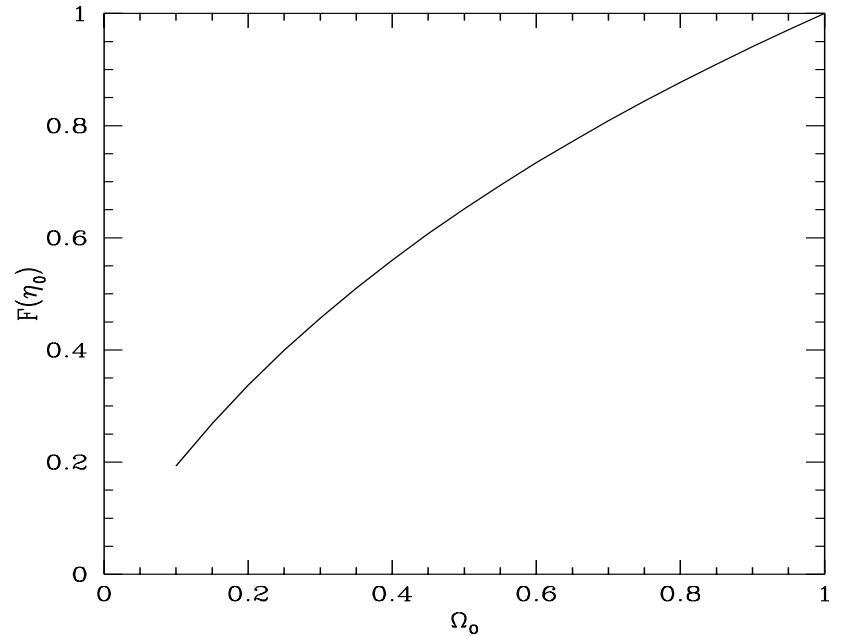

FIG. 3. The quantity $F(\eta_0)$ as a function of $\Omega$.

If $k \gg 1$ and $\eta \ll 1$, then the radial eigenfunctions $X_k^l(\eta)$ are approximated well by spherical Bessel functions, in which case

$$\tilde{\theta}_l(k) \simeq \frac{1}{3} j_l(k\eta_0) + 2 \int_{\eta_{ls}}^{\eta_0} \frac{dF}{d\eta}(\tilde{\eta}) j_l[k(\eta_0 - \eta)] d\tilde{\eta}, \qquad \text{for } l \gg l_{\rm curv}. \quad (5.2)$$

For $l \gg l_{\rm curv}$, microwave anisotropies come predominantly from perturbations with wavenumbers $k \gg 1$. The function $dF/d\eta$ is largest at late times when the Universe becomes curvature dominated. Therefore, at values of $\tilde{\eta}$ where the integrand in the line-of-sight term in Eq. (4.6) is appreciable, the argument of the radial eigenfunction is $\lesssim 1$, and should therefore be approximated well by a spherical Bessel function. On the other hand, if $\Omega$ is small, $\eta_0$ is not necessarily smaller than unity, so a spherical Bessel function does *not* provide a good approximation to the radial eigenfunction in the last-scattering term in Eq. (4.6). Still, for illustration, we include it in the analysis, although it should be kept in mind that the first term in Eq. (5.2) may be inaccurate.

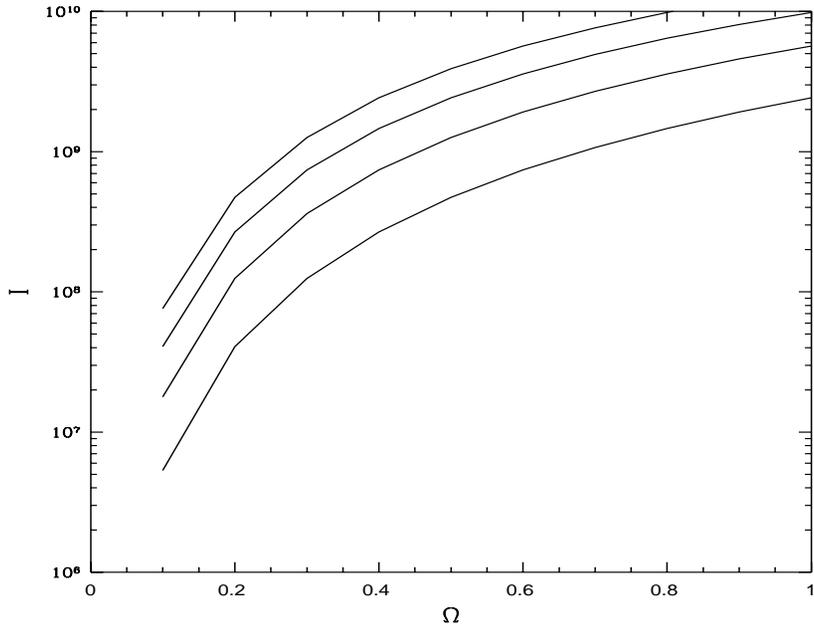

FIG. 4. The function $\mathcal{I}(\Omega,h)$ versus $\Omega$ for several values of $h$. The values of $h$, from the lowest to the highest curve, are $h = 0.4$, $h = 0.6$, $h = 0.8$, and $h = 1.0$.

Then, in the large-$l$ limit, the integral in Eq. (5.1) can be carried out analytically using the asymptotic relation (Kofman & Starobinsky 1986; Kamionkowski & Spergel 1993),

$$\int_0^\infty \frac{dk}{k} j_l(ak) j_l(bk) \sim \frac{\pi}{2l^3} a\delta(a-b) \qquad \text{as } l \to \infty, \tag{5.3}$$

where $\delta(a-b)$ is the Dirac $\delta$ function, so

$$\int [\tilde{\theta}_l(k)]^2 \frac{dk}{k} \simeq \frac{1}{18l(l+1)} \left[1 + \frac{g(\Omega)}{l}\right]. \tag{5.4}$$

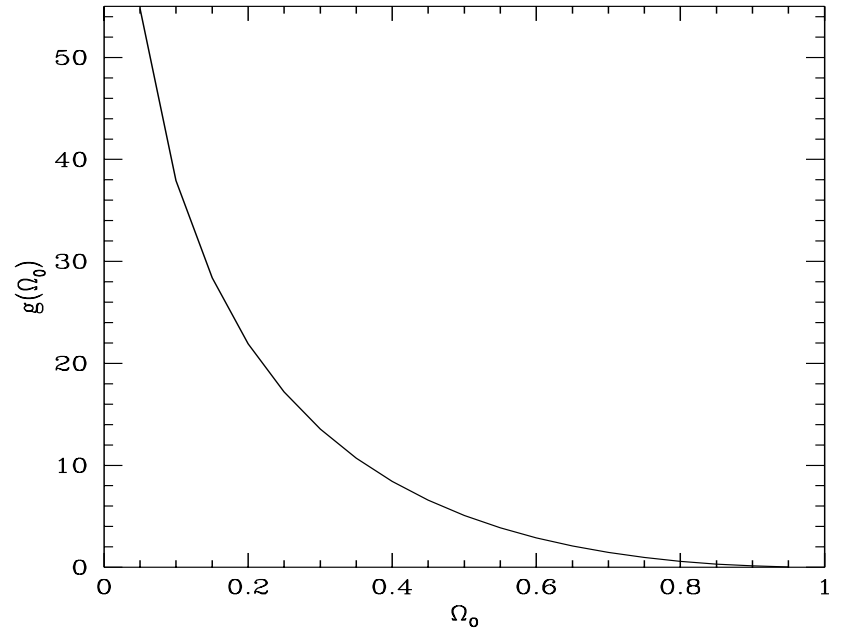

FIG. 5. The function $g(\Omega)$ vs $\Omega$.

Here, $g(\Omega)$ is a function of $\Omega$ given by

$$g(\Omega) = 36\pi \int_{\eta_{ls}}^{\eta_0} \left(\frac{dF}{d\eta}\right)^2 (\eta_0 - \eta) d\eta, \tag{5.5}$$

and is plotted in Fig. 5. The first term in Eq. (5.4) is the usual result describing the contribution from adiabatic density perturbations on the surface of last scattering. The second term in Eq. (5.4) is the contribution from the line-of-sight term. Note that the contribution to the CMB anisotropy from the line-of-sight term decreases relative to the last-scattering term as $l \to \infty$. In addition to the last-scattering and line-of-sight terms, there is also an interference term; however, it is easy to show, using Eq. (5.3), that the interference term becomes negligible as $l$ becomes large.

By comparing the analytic expressions, Eqs. (5.1) and (5.4), with numerical calculations of Eq. (4.8), we find that Eqs. (5.1) and (5.4) provide a fairly accurate approximation to the contribution from the line-of-sight term for $l \gtrsim 10$, but the result for the contribution from the last-scattering term is only good to about a factor of 2. Numerically, one finds that the last-scattering term is generally smaller than that given by the analytic approximation. This is most likely due to the fact that the spherical Bessel function does not provide a good approximation to the radial eigenfunctions at the surface of last scattering, as described above. The radial eigenfunctions $X_k^l(\eta)$ fall more rapidly with $\eta$ than do the spherical Bessel functions at distances larger than the curvature scale.

Eq. (5.1) and Fig. 4 make clear the basic $\Omega$ and $h$ dependence of the amplitude of the large-angle CMB anisotropies for a fixed normalization of the power spectrum on small scales. In addition, there is a weaker $\Omega$ dependence in the prefactor $[\Omega/F(\eta_0)]^2$. The $\Omega$ and $h$ dependences shown in Fig. 4 suggest that a model with $\Omega = 1$, $h = 0.5$, and no bias results in roughly the same quadrupole as a model with $\Omega = 0.3$, $h = 0.8$, and moderate biasing.

If the Universe is flat, but dominated by non-intersecting strings (or some other form of energy density that scales as $R^{-2}$), then the time dependence of the curvature perturbation is the same as that in an open Universe: Eq. (4.4). In this case, Eqs. (5.1) and (5.2) are exact, and Eqs. (5.4) and (5.5) provide an excellent approximation to the multipole moments for $l \gg 1$. The effects of curvature are the only source of difference between the large-angle CMB anisotropies in an open universe and the CMB anisotropies in a string-dominated flat universe. Therefore, in the next Section we will compare these results to isolate the effect of curvature. In addition, these results can also be applied to other flat-space models that are not matter dominated, such as a cosmological-constant Universe (Kofman & Starobinsky 1986; Kamionkowski & Spergel 1993).

## 6. NUMERICAL RESULTS

As pointed out in the Introduction, multipole moments $l \lesssim l_{\text{curv}}$ probe scales larger than the curvature scale. We will compare our results for the

17

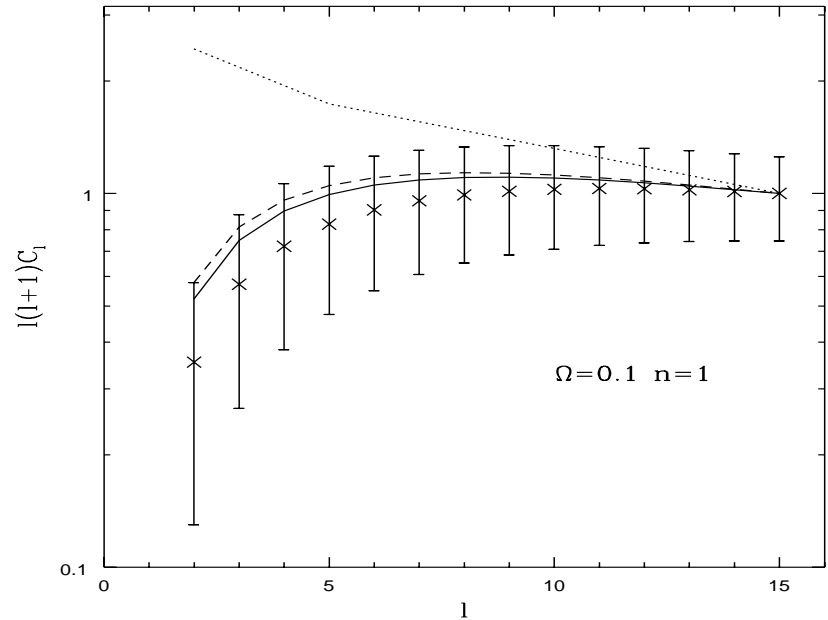

FIG. 6. The predicted spectrum of CMB multipole moments, with arbitrary normalization, for adiabatic perturbations with an $n = 1$ primordial spectrum that is a power law in volume. The error bars are the theoretical uncertainties due to cosmic variance. Also shown are the results for primordial spectra that are $n = 1$ power laws in wavenumber $k$ (solid curve) and eigenvalue $q = (k^2 + 1)^{1/2}$ (broken curve). The dotted curve is the spectrum for a string-dominated flat universe with $\Omega = 0.1$ contributed by non-relativistic matter.

multipole moments with the results of the COBE DMR experiment (Wright 1993).

In addition to the observational errors, cosmic variance must also be considered when comparing calculations of the multipole moments with the observational results. We make the simplest and most plausible assumption that the multipole moments have a Gaussian distribution, so given the power spectrum, the $1\sigma$ uncertainty in the theoretical prediction for each $C_l$ is $\sqrt{2/(2l+1)}C_l$.

18

In Fig. 6, we plot $l(l+1)C_l$ against the multipole moment $l$ for an $n = 1$ volume scaling in an $\Omega = 0.1$ universe with arbitrary normalization. The uncertainties plotted are those due to cosmic variance. Standard CDM would predict a flat curve. In addition, we also plot the multipole moments expected with primordial power spectra that are $n = 1$ power laws in wavenumber $k$ and eigenvalue $q$. As expected from from Fig. 1, these scalings produce larger CMB anisotropies on large scales. Even so, note that for all three *ansatzes* for the primordial spectrum (which look quite different on large scales; c.f. Fig. 1), CMB anisotropies on scales larger than the curvature scale are suppressed relative to those on scales smaller than the curvature scale.

Also plotted (dotted curve) in Fig. 6 are the CMB moments for a string-dominated flat universe with $\Omega = 0.1$ contributed by nonrelativistic matter. As explained in the previous Section, curvature is the only source of difference between the CMB anisotropies in an open universe and those in a string-dominated Universe. The dotted curve clearly illustrates that curvature, not time-dependence of the potential, is responsible for the large-angle suppression in an open Universe.

In Fig. 7 and Fig. 8, we plot the integrand in Eq. (4.8) for an $n = 1$ volume scaling with $\Omega = 0.1$. Shown are the total integrand (solid curve), the integrand which would be obtained by including only the line of sight integral in $\tilde{\theta}_l$ (long-dash curve), and the integrand which would be obtained by including only the last-scattering term in $\tilde{\theta}_l$ (short-dash curve). In Fig. 7 the contributions to the $l = 2$ multipole moment are shown, and in Fig. 8 the contributions to the $l = 15$ multipole moment are shown. The oscillatory behavior of the last-scattering term comes from the $k$ dependence of the function $X_k^l(\eta_0 - \eta_{ls})$ in Eq. (4.6). The line-of-sight term does not oscillate as a function of $k$ since the contribution to a given value of $k$ results from an integral over conformal time, $\eta$. As shown in Fig. 7 and Fig. 8, in a low-$\Omega$ model, the dominant contribution to large-angle anisotropies comes from the line-of-sight integral in Eq. (4.6), in agreement with the analysis in the previous Section.

One of the primary motivations for considering low-$\Omega$ models are recent measurements which find $h \simeq 0.8$, clearly in conflict with a flat universe, which favors $h = 0.5$. Therefore, it is natural to consider high values of $h$ if we consider low values of $\Omega$. First we consider the standard CDM model ($\Omega = 1$) and take $h = 0.5$. By fitting the COBE (Wright 1993) and Tenerife (Davies et al. 1992; Watson et al. 1992; White, Silk, & Scott 1993) measurements to a Harrison-Zel'dovich power spectrum, and using the Bond-Efstathiou transfer function, Eq. (3.14), we find that standard CDM fits the observed level of galaxy fluctuations on small scales (i.e. $\sigma_8 \sim 1$) without any biasing of galaxies relative to mass: $\sigma_8^{\rm mass} = 0.9 \pm 0.3$. If we consider $\Omega = 0.3$ and fit our $n = 1$ volume-scaling power spectrum with $h = 0.8$, we find that $\sigma_8^{\rm mass} = 0.55 \pm 0.15$. An $\Omega = 0.1$ Universe with $h = 0.8$ predicts too low an amplitude of mass fluctuations, $\sigma_8^{mass} = 0.15 \pm 0.04$. Keep in mind, however, that there is a significant statistical uncertainty in the value of $\sigma_8^{mass}$ expected from normalization to COBE and Tenerife; the $1\sigma$ errors in $\sigma_8^{mass}$ are roughly 20-30%. In Fig. 9 we plot the multipole moments obtained from the COBE first-year data (Wright 1993), a point at $l = 20$ from the Tenerife data (Davies et al. 1992; Watson et al. 1992; White, Silk, & Scott 1993), as well as the best fits for $\Omega = 0.1$, (dashed curve), $\Omega = 0.3$ (solid curve), and $\Omega = 1$ (dotted curve).

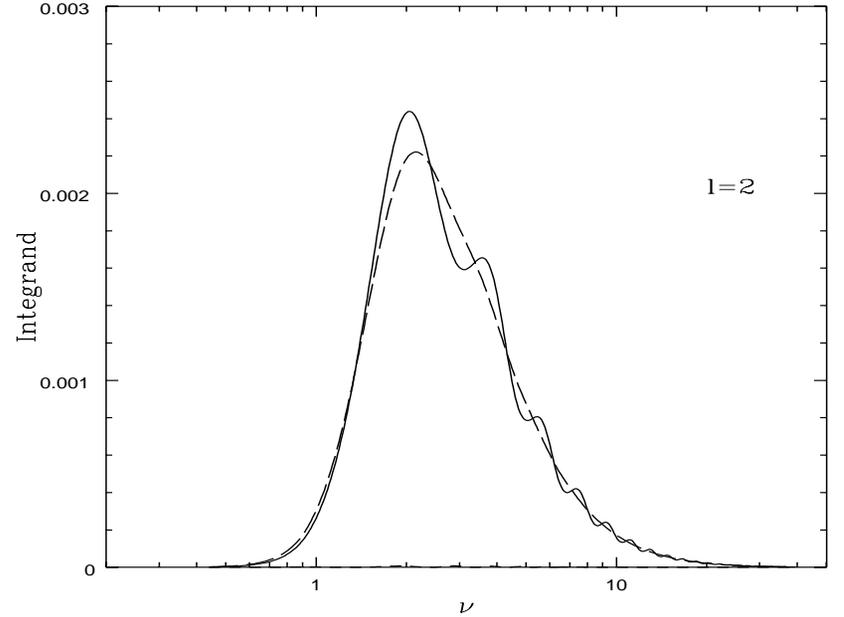

FIG. 7. The integrand of the multipole moment $l = 2$ for a volume scaling with $\Omega = 0.1$. Shown are the total integrand (solid curve), the integrand which would be obtained by including only the line of sight term in $\tilde{\theta}_l$ (dash curve), and that which would be obtained by including only the last-scattering term in $\tilde{\theta}_l$ (short-dash curve).

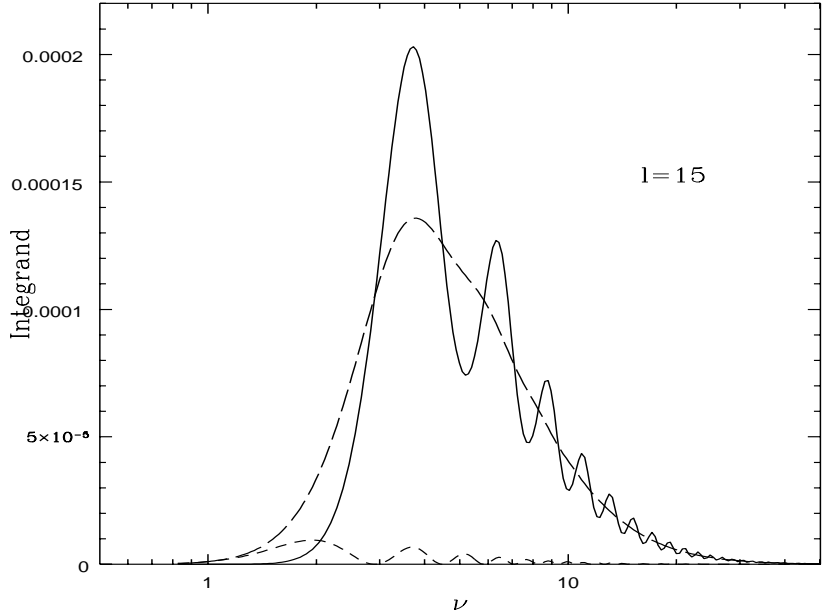

FIG. 8. Same as Fig. 7 for $l = 15$.

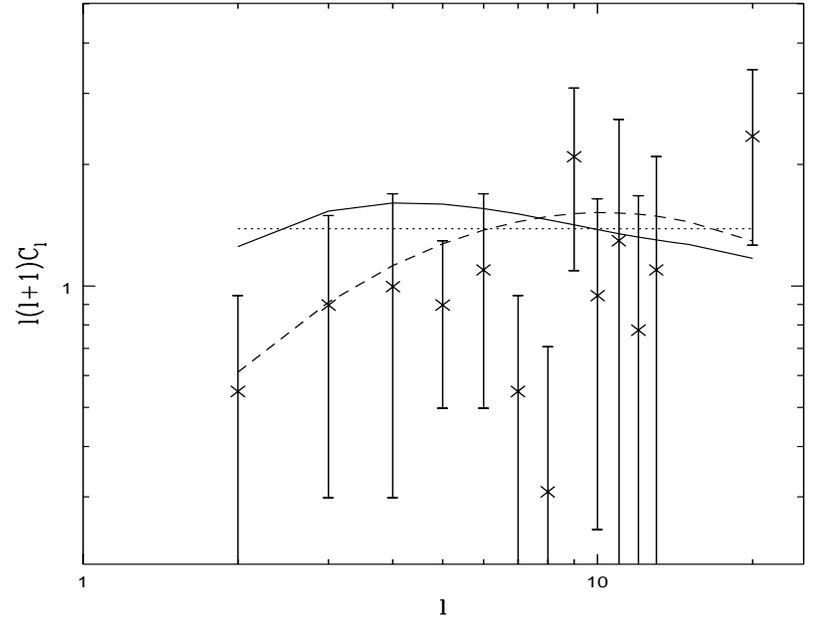

FIG. 9. The COBE multipole moments from Wright (1993). The solid (dashed) curve is the best fit for an $\Omega = 0.3$ (0.1) model with a volume scaling. The dotted curve is the best fit for an $\Omega = 1$ scale-invariant spectrum. Also plotted (the $l = 20$ moment) is an estimate of the Tenerife result (Davies et al. 1992; Watson et al. 1992; White, Silk, & Scott 1993).

As pointed out in the previous Section, although the magnitude of the predicted quadrupole moment increases as $\Omega$ is decreased with all other parameters fixed, it also decreases as $h$ is increased. Upon examining Eqs. (4.8) and (4.6), we find that the *shape* of the spectrum on large angular scales does not depend on the Hubble parameter. The Hubble-parameter dependence of the *amplitude* of the spectrum enters through the function $\mathcal{I}(\Omega, h)$ as shown in Eq. (5.1). Therefore, for a given $\Omega$, it is simple to scale the value of $\sigma_8^{mass}$ required to fit COBE with power on small scales with $h$. The results are shown in Fig. 10; values of $\Omega = 1$, 0.3, and 0.1 are represented by the $\times$'s, triangles, and squares, respectively. Thus if we are willing to accept that $\sigma_8^{mass} \gtrsim 0.5$, for example, an $\Omega = 0.3$ universe is consistent within $3\sigma$ with COBE observations, even with a fairly low value of $h$; however, a value of $\Omega = 0.1$ with an $n = 1$ primordial spectrum is clearly inconsistent. Similar conclusions should apply for power spectra that scale with length or eigenvalue of the Laplacian.

In addition, since the shape of the spectrum on COBE scales is independent of the Hubble parameter, we can (at least in principle) use the measured data points to discriminate between the various values of $\Omega$, assuming the given power-law spectra. This is done by evaluating the value of the reduced $\chi^2$ obtained from fitting the models to the observations. With the current data, the reduced $\chi^2$ for all three values of $\Omega$ are much smaller than unity, so the current COBE data alone cannot be used to discriminate between the models. As discussed above, due to cosmic variance, this is likely to remain the case even with improved observations. However, with more information on smaller angular scales, we may be able to use CMB anisotropies to distinguish between the various models.

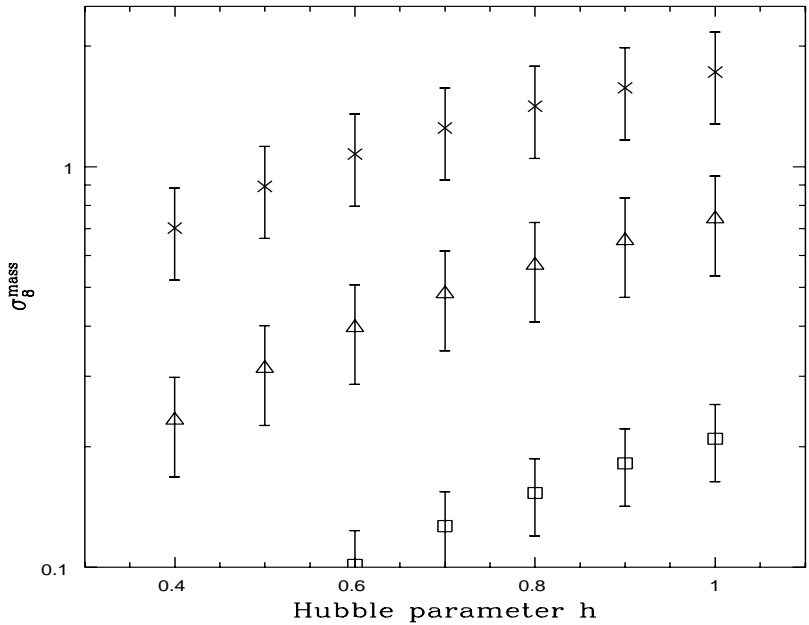

FIG. 10. The amplitude, $\sigma_8^{\text{mass}}$, of mass fluctuations obtained by normalizing the power spectrum to COBE and smoothing the density field with an $8h^{-1}$ Mpc top hat. Values of $\Omega = 1$, 0.3, and 0.1 are represented by the ×'s, triangles, and squares, respectively.

In Fig. 11, we plot contours of $\sigma_8^{\text{mass}}$ obtained by normalizing to COBE, as well as contours of age of the Universe, $t_U$, in the $\Omega$-$h$ plane. The heavy solid curve is the contour of $\sigma_8^{\text{mass}} = 1$; the upper and lower lighter solid curves are contours of $\sigma_8^{\text{mass}} = 0.7$ and 0.5, respectively. The upper and lower broken curves are contours of age of the universe of 10 and 13 Gyrs respectively. The shaded region is that where $0.2 \leq \Omega h \leq 0.3$, as suggested by the observed power on large scales (Vogeley et al. 1992; Fisher et al. 1993a; Baugh & Efstathiou 1993; Peacock & Dodds 1993).

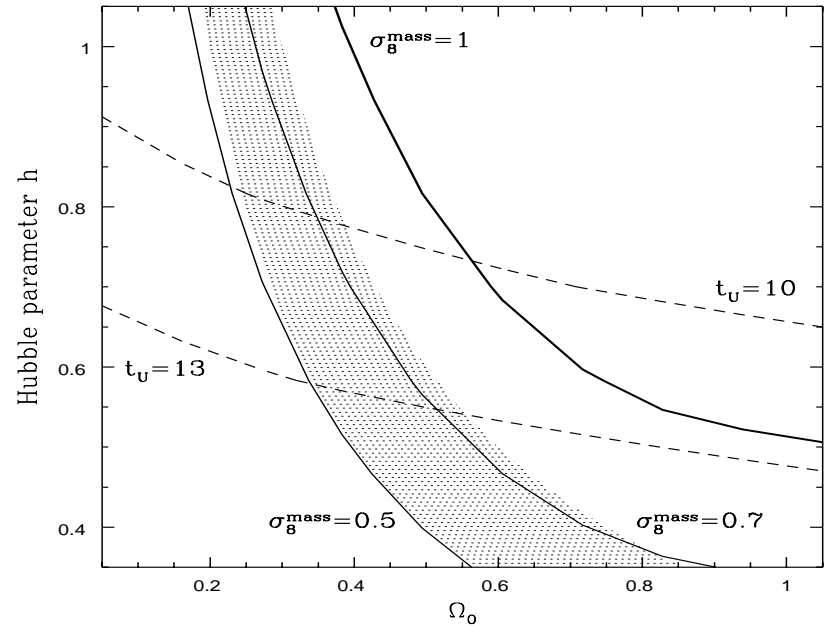

FIG. 11. Contour plot of $\sigma_8^{\text{mass}}$ obtained from COBE normalization in the $\Omega$-$h$ plane. The heavy solid curve is the contour of $\sigma_8^{\text{mass}} = 1$; the upper and lower lighter solid curves are contours of $\sigma_8^{\text{mass}} = 0.7$ and 0.5, respectively. The upper and lower broken curves are contours of age of the universe of 10 and 13 Gyrs respectively. The shaded region is that where $0.2 \leq \Omega h \leq 0.3$, as suggested by the observed power on large scales.

Conservative lower limits on the ages of globular clusters constrain the parameter space to the region below the lower broken curve in Fig. 11. This Figure shows that if we are willing to accept values of $\sigma_8^{\text{mass}} \gtrsim 0.5$, and take $h \gtrsim 0.4$, both the age of the Universe and the observed power on large scales can be fit with values of $0.4 \lesssim \Omega \lesssim 0.8$; this requires a Hubble constant $\lesssim 0.6$. We should also point out that the COBE normalized values of $\sigma_8^{mass}$ currently have error bars of about 20-30% (c.f. Fig. 10); therefore, even if we are willing to accept a more conservative range of values for the mass fluctuation, e.g. $\sigma_8^{mass} \gtrsim 0.7$, the region bounded by the $\sigma_8^{mass} = 0.5$ curve is still consistent with $\sigma_8^{mass} > 0.7$ at the $2\sigma$ level.

Large-scale correlations come with significant uncertainties, so the range of values of $\Omega h$ may admittedly be larger than shown here; even so, $\Omega = 1$ is difficult to accommodate unless $h$ is unexpectedly small.

If we disregard the globular cluster ages and take the more conservative bound of 10 Gyr on the age of the Universe, then larger values of the Hubble constant ($0.6 \lesssim h \lesssim 0.8$) can be accommodated with $\sigma_8^{mass} \gtrsim 0.5$ and $0.2 \lesssim \Omega h \lesssim 0,3$ with $0.2 \lesssim \Omega \lesssim 0.6$. On the other hand, if the Universe is flat (and matter dominated) higher values of the Hubble constant cannot be accommodated.

## 7. SUMMARY AND CONCLUSION

After the announcement of the detection of CMB fluctuations by the DMR team, some cosmologists interpreted the detection as evidence that the Universe is flat and that the fluctuations reflect variations in the potential at the surface of last scatter at $z \sim 1300$. Neither of these statements need be true.

We have found that the amplitude and multipole spectrum detected by COBE is compatible with a scale-invariant spectrum of fluctuations in an open universe. We have explored various possible ways of extending our notion of scale invariance to scales comparable to and larger than the curvature scale. Despite the fact that the density power spectrum is very sensitive to how scale invariance is extrapolated to large scales, the predicted level of CMB fluctuations is much less sensitive to our assumptions (compare Fig. 1 and Fig. 6).

A scale-invariant spectrum of density fluctuations normalized to COBE's detection predicts mass fluctuations on the $8h^{-1}$ Mpc scale: $\sigma_8^{mass} \simeq 0.9(h/0.5)$ in a flat $\Omega = 1$ universe, $\sigma_8^{mass} \simeq 0.6(h/0.8)^{1.2}$ in an $\Omega = 0.3$ universe, and $\sigma_8^{mass} \simeq 0.2 h^{1.5}$ in an $\Omega = 0.1$ universe. These predictions should be compared with the observed level of galaxy fluctuations on this scale: $\sigma_8 = 0.83$ in the APM survey (Baugh & Efstathiou 1993) and $\sigma_8 \simeq 1$ determined from the 14.5-mag CfA survey (Davis & Peebles 1983). Thus, for small values of the Hubble constant, $h < 0.6$, and the flat-universe models appear compatible with the observed level of COBE fluctuations, while for larger values of the Hubble constant, the low-$\Omega$ models appears favored. Observations of galaxy fluctuations on scales larger than $8h^{-1}$ Mpc are better fit by a scale-invariant spectrum with $\Omega h \sim 0.25$ (Vogeley et al. 1992; Fisher et al. 1993a; Baugh & Efstathiou 1993; Peacock & Dodds 1993). Numerical simulations of clusters (Bahcall & Cen 1992) when compared to observations find that cluster properties are best fit by models with $\Omega \sim 0.2 - 0.3, h \sim 0.75$ and $\sigma_8^{mass} \sim 0.7 - 1$, which is consistent with our inferred COBE normalization.

Unlike in a flat matter-dominated model, where the low-multipole fluctuations reflect variations in the gravitational potential at the surface of last scatter, the dominant source of CMB fluctuations on large angular scales in the open models examined here are variations in the gravitational potential at $z \sim \Omega^{-1}$. Thus, COBE need not be detecting the gravitational potential at $z \sim 1300$.

Unfortunately, it will be difficult for COBE to distinguish between low-$\Omega$ and flat models as the large cosmic variance of the low multipoles can erase the difference in multipole spectra. While the low multipoles are suppressed in the models with curvature fluctuations explored in this paper, the effect is not as dramatic as in isocurvature models which predict larger suppressions of the low multipoles (Peebles 1987a,b; Spergel & Pen 1993). In addition, there is also the possibility that tensor modes (i.e. gravitational waves) may provide a significant contribution to the microwave anisotropies on large angular scales (Bond et al. 1993; Caldwell 1993).

We suspect that observations of higher multipoles moments may be better able to distinguish between low-$\Omega$ and flat models. If the Universe was not reionized at $z \sim 50 - 100$ by early object formation, then the predicted microwave fluctuations near the Doppler peak at $l \sim 200\Omega^{-1}$ exceed current observational limits at these small angular scales (Bardeen, Bond, & Efstathiou 1987). On the other hand, low-$\Omega$ models predict higher amplitude fluctuations at large z, thus, it is not unreasonable to imagine that early object formation could reionize the background and suppress fluctuations on angular scales smaller than $\Omega(1+z_{ls})^{-1/2}$, where $z_{ls}$ is the redshift at which the optical depth was unity. This suggests that observations at the degree scale should be able to distinguish between open and flat models. We will address this question in a subsequent paper (Kamionkowski & Spergel 1993).

## 8. ACKNOWLEDGMENTS

MK thanks R. Caldwell, J. Frieman, A. Liddle, D. Scott, J. Silk, A. Stebbins, N. Sugiyama, M. Turner, and M. White, for useful discussions and suggestions. DNS thanks A. Vilenkin for noting the connection between open models and string-dominated universes. MK acknowledges the hospitality of the Center for Particle Astrophysics and the Aspen Center for Physics, where part of this work was completed. MK was supported in part by the Texas National Research Laboratory Commission and by the U.S. Department of Energy under contract DEFG02-90-ER40542. DNS was supported in part by NSF grant AST88-58145 and NASA grant NAGW-2448.